# Tunable Anisotropic Thermal Transport in Super-Aligned Carbon Nanotube Films

*Wei Yu, [a][‡] Xinpeng Zhao, [b][‡] Puqing Jiang, [c][‡] Changhong Liu, [a][*] and Ronggui Yang[c][*]*

[a]Tsinghua-Foxconn Nanotechnology Research Center and Department of Physics, Tsinghua University, Beijing 100084, China

[b]Department of Mechanical Engineering, University of Colorado, Boulder, Colorado 80309, USA

[c]School of Energy and Power Engineering, Huazhong University of Science and Technology, Wuhan 430074, China

**Abstract**

Super-aligned carbon nanotube (CNT) films have intriguing anisotropic thermal transport properties due to the anisotropic nature of individual nanotubes and the important role of nanotube alignment. However, the relationship between the alignment and the anisotropic thermal conductivities was not well understood due to the challenges in both the preparation of high-quality super-aligned CNT film samples and the thermal characterization of such highly anisotropic and porous thin films. Here, super-aligned CNT films with different alignment configurations are designed and their anisotropic thermal conductivities are measured using time-domain thermoreflectance (TDTR) with an elliptical-beam approach. The results suggest that the alignment configuration could tune the cross-plane thermal conductivity $k_z$ from 6.4 to 1.5 W/mK and the in-plane anisotropic ratio from 1.2 to 13.5. This work confirms the important role of CNT alignment in tuning the thermal transport properties of super-aligned CNT films and provides an efficient way to design thermally anisotropic films for thermal management.

**Introduction**

Carbon nanotube (CNT)-based materials have attracted great interest in a range of emerging thermal management technologies, from heat dissipation of high power electronics and photonics[1-3] to personal thermal management,[4, 5] because of their potentially high thermal conductivity, excellent mechanical strength/flexibility, and chemical stability. Previous research showed that the thermal conductivity of a single CNT along the axial direction could reach as high as ~1000 W/mK,[6] while the thermal conductivity of a three dimensional (3D) random CNT aerogel was only ~0.06 W/mK,[7] differing by five orders of magnitude from thermal superconductor to thermal super-insulator. This immense contrast suggests that not only the intrinsic thermal properties of individual CNTs but also the alignment and the contact resistance between the tubes play important roles in the thermal conductivity of CNT-based materials.[8, 9] Highly anisotropic thermal transport is expected in super-aligned CNT films[10-12] and is to be highly dependent on the alignment. With highly tunable anisotropic thermal transport properties, these super-aligned CNT films could be better utilized for directional control of heat flow, which is an important application of thermal management. Therefore, understanding how the alignment configuration affects the anisotropic thermal transport in super-aligned CNT films is vitally important to the application of those CNT-based materials.

However, the anisotropic thermal conductivities of super-aligned CNT films are not easy to measure due to many constraints, including the micrometer thicknesses, the porous internal structures, and the 3D anisotropic nature of thermal transport in these super-aligned CNT films. Our previous work has shown that the time-domain thermoreflectance (TDTR) measurement technique with a highly elliptical pump beam could be used to probe the 3D anisotropic thermal conductivity tensor.[13, 14] However, the porous nature of the super-aligned CNT films cannot meet the requirement of an optically flat and reflective surface for TDTR measurements.

In this work, high-quality super-aligned CNT films with different alignment configurations were fabricated, and their 3D anisotropic thermal conductivity tensors were measured using the elliptical-beam TDTR. In particular, to measure the porous film, the samples were sandwiched to meet the needs of TDTR measurements. The results indicate that both the in-plane and the cross-plane thermal conductivity can be tuned by different alignment configurations of the CNT bundles. The cross-plane thermal conductivity $k_z$ could be tuned from 6.4 to 1.5 W/mK when the stacking angle $\theta$ (the angle between the alignment directions of two adjacent CNT layers) increases from 0° to 90°. The in-plane thermal conductivities are also highly directional-dependent and easily subject to the alignment. The in-plane anisotropic ratios are 13.5, 2.8 and 1.2 for CNT films with $\theta = 0°$, 90°, and 45°, respectively. The experiments suggest that the highly tunable 3D anisotropic thermal transport of super-aligned CNT films originates from both the anisotropic nature of individual carbon nanotubes and the inter-tube contact resistance. This work not only has a great significance in understanding the relationship between thermal conductivity and CNT alignment but also provides a solution for the thermal conductivity measurement of porous thin films with high thermal anisotropy.

**Results and discussion**

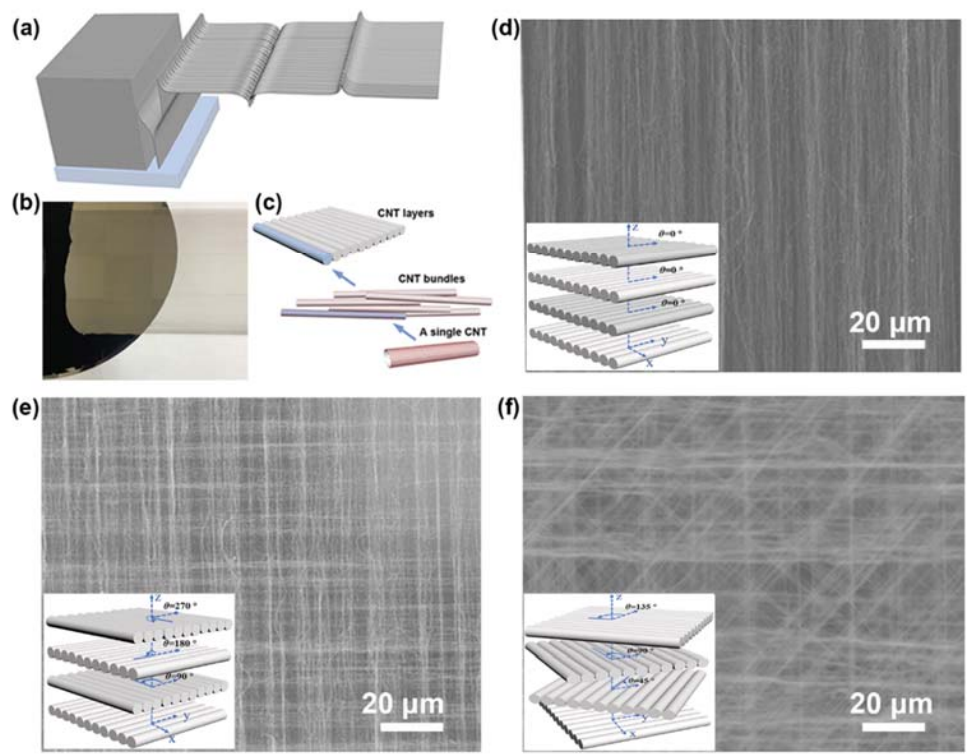

*Figure 1. (a) Schematic of CNT films drawn from super-aligned CNT arrays. (b) Optical image showing one layer of the CNT film being drawn from the super-aligned CNT arrays. (c) Schematic of the structure of CNT layers: the CNT layers consist of super-aligned CNT bundles that are composed of many CNTs bonded together due to van der Waals interaction. (d-f) SEM images of CNT films with stacking alignment angle of (d) θ = 0°, (e) θ = 90° and (f) θ = 45°. The inserts in (d), (e), and (f) are their schematic diagrams.*

The super-aligned carbon nanotube (CNT) arrays were first synthesized on silicon substrates by chemical vapor deposition. Most of the CNTs are multiwalled tubes with a radius of 10-20 nm and a height of ~300 $\mu$m.[10, 11, 15, 16] Due to the high nucleation density and narrow size distribution of catalysis, the alignment and purity of the super-aligned CNT arrays are much better than the common vertically aligned CNT arrays.[10, 16, 17] Continuous CNT films with thicknesses of tens of nanometers were then directly drawn from the CNT arrays as shown in Figure 1(a, b). The CNT films are aggregated by CNT bundles which are composed of many CNTs due to the van der Waals interaction (Figure 1c). By stacking the monolayer films layer by layer with different stacking angles $\theta$, the super-aligned multilayer CNT films were obtained, with Figure 1d, e, and f showing films with different stacking angles of $\theta$= 0°, 90°, and 45°, respectively. With the CNT bundles distribute orderly with certain stacking angles, these unique structures result in high-purity, high-quality, and quasi-unidirectionally aligned CNT arrays. Besides those shown in Figure 1, other CNT films with stacking angles $\theta = 30°, 60°$ were also fabricated and measured. It is expected that the thermal transport in the CNT films is highly anisotropic, closely related to the directional arrangement of CNT bundles. For comparison, the randomly aligned CNT films are also fabricated and measured to compare with the super-aligned films (Supporting information, Figure S7a).

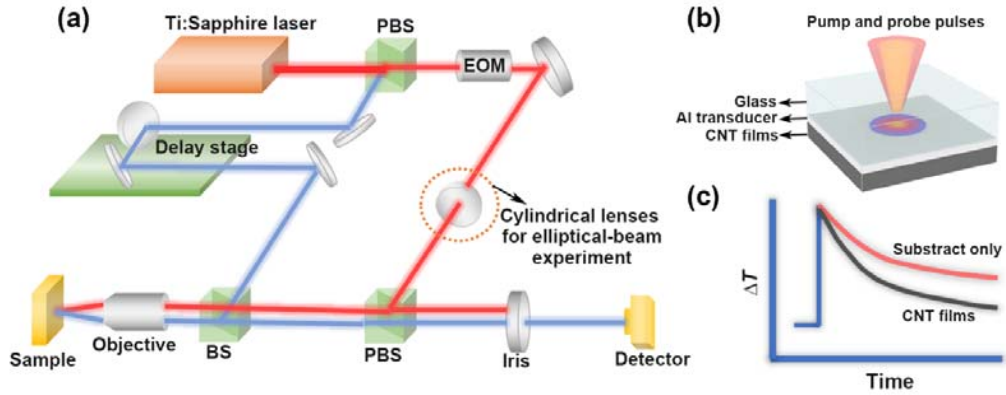

*Figure 2. (a) Schematic of the TDTR setup used for measuring the three-dimensional anisotropic thermal conductivity of CNT films. (b) Schematic of the sandwich sample configuration used for TDTR measurement. Unlike the conventional structures, the Al transducer film with thickness ~100 nm is sandwiched between a transparent glass slide and the CNT films to be measured. The pump and probe beam need to penetrate through the glass slide before impinging on the Al transducer film. (c) The relative cooling rates for the sandwiched sample (the Al transducer sandwiched between the CNT film and the glass substrate) and the substrate sample (the glass substrate coated with the Al transducer).*

The anisotropic thermal conductivities of CNT films were measured using a femtosecond laser-based TDTR system, as shown in Figure 2a. More details of the system could be found in our previous work.[18, 19] Unlike TDTR measurements of optically flat samples, the aluminum (Al) transducer layer cannot be directly deposited onto the surface of the CNT films due to the internal porosity. To solve this problem, a special sandwiched structure as shown in Figure 2b was designed, where the Al transducer layer with a thickness of ~100 nm was sandwiched between the CNT film and a transparent glass slide. Both the pump (heating) and probe (sensing) beam penetrate through the transparent glass before impinging on the Al transducer film. The temperature change was monitored by the probe laser through the temperature-dependent change in the optical reflectance of the transducer. The relative cooling rates for the case of bi-directional heat flow from the Al film to both the CNT sample and the glass would be different from the case of uni-directional heat flow from the Al film to the glass, as illustrated in Figure 2c. The biggest challenge for this sandwiched sample configuration to work is the weak bonding between the CNT/Al interface through the van der Waals force, resulting in an ultra-low thermal conductance ($G$) that could make the measurements insensitive to the thermal conductivity of the CNT samples.[3, 20-22] We find that the sensitivity to the cross-plane thermal conductivity $k_z$ could be lower than 0.05 when $G$=1.0 MW/(m²K), as shown in Figure S2a. Here the sensitivity coefficient is defined as $S_\gamma = \frac{\gamma}{R}\frac{\partial R}{\partial \gamma}$, where $\gamma$ is the parameter that we are interested in and $R = -V_{in}/V_{out}$ is the detected signal.[23, 24] If $G$ is higher than 10 MW/(m²K), the sensitivity to $k_z$ would increase to be larger than 0.085, making the measurements of $k_z$ possible. Using a large laser spot size of 40 $\mu$m and a relatively high modulation frequency $f$ of 2-10 MHz could suppress the sensitivity to in-plane thermal conductivity while maintaining a high sensitivity to the cross-plane thermal conductivity (Figure S2b). To improve the thermal conductance of the CNT/Al interface, isopropanol was sprayed onto the CNT film, which is then placed on the Al surface, and a pressure of ~ 0.65 MPa was applied to the sandwich structure for 24 hours. Here it is noted that although the cross-plane thermal conductivity $k_z$ could be measured after the $G$ was improved, the $G$ could not be determined accurately due to its very low sensitivity ($S_G < 0.05$, as shown in Figure S2b). The indeterminable $G$, however, does not cause any barrier for the current study as it does not affect our measurements of $k_z$ of the CNT films, which is the aim of this study. Separate TDTR measurements reveal the thermal conductivity of glass as 1.35±0.13 W/mK and the interface thermal conductance ($G_1$) of glass/Al as 300±30 MW/(m$^2$K). The thickness of the Al transducer was determined to be 103±3 nm by picosecond acoustics. The specific heat capacity of CNTs is taken as 0.7 J/(gK) from the literature.[25, 26] The thicknesses of the CNT films were determined to be ~ 6 $\mu$m using both a digital micrometer and a profiler. Such thick CNT films are considered "thermally thick" in our experiments, considering that the largest cross-plane thermal penetration depth in our TDTR measurements is <2 $\mu$m. The density of the CNT film was determined by dividing its mass by its volume as 0.22±0.05 g/cm$^3$. To validate TDTR measurements of the sandwich-structured samples, a 2.0-μm-thick amorphous silica film was deposited on the Al side of an Al-coated glass by sputtering deposition, and its thermal conductivity was determined as 1.32±0.6 W/mK (Figure S1a), which is in excellent agreement with the literature.[27]

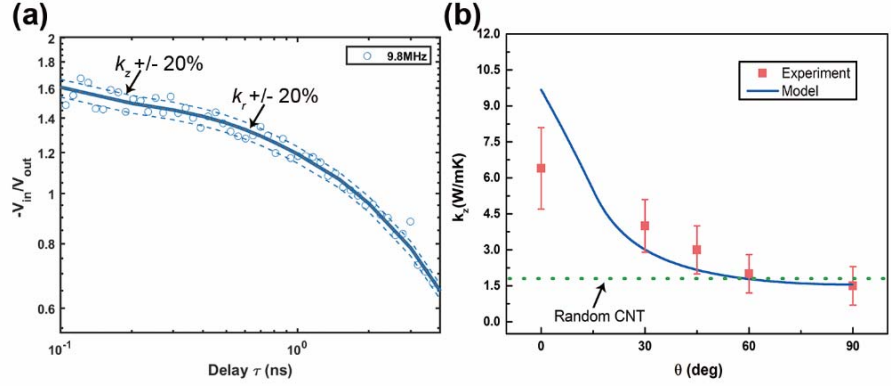

*Figure 3. (a) TDTR signals and their fitting curves for the cross-plane thermal conductivity of CNT films with stacking angle θ=0°, in which the spot radius of w=40 μm and the modulation frequency of 9.8 MHz were used. The curves of 20% bounds to the best-fitted thermal conductivity values are included to demonstrate the sensitivity of the signals. (b) The cross-plane thermal conductivity of CNT films with different stacking angle θ. Symbols represent measured results using the TDTR method, and the solid line is the predicted results using the thermal network model. The dashed line indicates the cross-plane thermal conductivity of the random CNT film from TDTR measurements.*

The cross-plane thermal conductivity $k_z$ of the CNT films could be first determined independently from measurements using a large spot radius of 40 $\mu$m and a high frequency of 9.8 MHz (see sensitivity analysis in Figure S2b). Figure 3a shows an example of the TDTR signals $R = -V_{in}/V_{out}$ of the CNT film with stacking angle $\theta = 0°$ and their fitting curve as a function of the delay time between the pump and the probe beam. The corresponding cross-plane thermal conductivity was determined to be $k_z$ = 6.4 ± 1.7 W/mK. To explore the influence of the stacking angle $\theta$ on $k_z$, $k_z$ of the aligned CNT films with $\theta$ = 30°, 45°, 60°, and 90° were measured (Figure 4b). It is found that $k_z$ decreases as the stacking angle $\theta$ increases from 0° to 90°, with the 90° aligned CNT film having the lowest $k_z$ (1.5±0.8 W/mK), even slightly lower than that of the random CNT film (1.8±0.9 W/mK) (Figure S7d). The reason for this stacking angle-dependence of $k_z$ is that the thermal resistance in the $z$-direction is dominated by both the thermal resistance of CNT bundles $R_{z,CNT}$ and the inter-bundle contact resistance $R_{z,c}$ (Figure S5a and S5b). Assuming the same length $L_0$ and diameter $D$ for all the CNT bundles, the cross-plane thermal conductivity of the aligned CNT films could be written as

$$k_z = \frac{D}{R_{z,CNT}+R_{z,c}} = nA_z k_\perp \frac{Bi_z}{1+Bi_z}, \qquad (1)$$

Where the Biot number $Bi_z = h_z D/A_z k_\perp$ characterizes the ratio of thermal resistance of the CNT bundles and the interface thermal resistance, $A_z$ is the cross-section areas of a single CNT bundle normal to the z-direction, $k_\perp$ is the thermal conductivities of CNT bundles perpendicular to the axial direction, and $h_z = h_0 A$ with $h_0$ being the interface thermal conductance per unit area and $A$ being the contact area between the CNT bundles in the adjacent layers. The contact area would decrease as the stacking angle $\theta$ increases from 0° to 90° (Supporting Information, Equation S28), resulting in increased inter-tube contact resistance (decreased $Bi_z$); as a result, the $k_z$ of super-aligned CNT films decreases with $\theta$. The $k_z$ of super-aligned CNT films predicted by Equation (1) as a function of the stacking angle $\theta$ is plotted as the curve in Figure 3b, which agrees qualitatively well with our measurements.

With the cross-plane thermal conductivity $k_z$ determined, we can then measure the anisotropic in-plane thermal conductivities of the CNT films using the elliptical-beam TDTR approach. Using an elliptical laser spot of $w_a = 16\ \mu m$ and $w_b = 4\ \mu m$ and a low modulation frequency of 1.06 MHz, the sensitivity to the in-plane thermal conductivity along the long axis direction $k_a$ can be successfully suppressed (with $S_{k_a} < 0.065$) while the sensitivity to the in-plane thermal conductivity along the short axis direction $k_b$ remains high ($S_{k_b} > 0.12$, Figure S2c). To generate an elliptical pump spot on the sample surface, a pair of cylindrical lenses were inserted in the pump line, as shown in Figure 2a, whereas the probe spot remains circular. Details of the elliptical-beam TDTR approach can be found in Ref. 18. The directional dependence of in-plane thermal conductivity could be obtained by rotating the CNT samples. For validation, the thermal conductivity of sapphire was measured as 33.7±4 W/mK using this elliptical beam method (Figure S1c).

Figure 4 shows the in-plane thermal conductivities of three CNT films with different stacking angles of $\theta$ = 0°, 90°, and 45°, respectively, for the different in-plane directions $\alpha$. For the parallel-aligned CNT film (with $\theta$ = 0°), the largest in-plane thermal conductivity is determined as 90.5±18.1 W/mK along the $\alpha$ = 0° direction, and the smallest is determined as 6.7±2 W/mK along the $\alpha$ = 90° direction (Figure 4a), with the anisotropic ratio reaching as high as 13.5. For the cross-aligned CNT film (with $\theta$ =90°), $k_r$ changes with a period of 90° due to the in-plane geometric symmetry. The highest thermal conductivity was found as ~ 78.1 W/mK along both the $\alpha$ = 0° and 90° directions, and the lowest thermal conductivity was found as ~ 28.3 W/mK along the $\alpha$ = 45° direction. The anisotropic ratio reduces to 2.8, as shown in Figure 4b. For the CNT film with $\theta$ =45°, the anisotropic ratio further reduces to ~ 1.2, showing an approximately homogenous in-plane thermal conductivity, as shown in Figure. 4c. These results indicate that the alignment direction of the CNTs has a great impact on the in-plane thermal conductivity.

Theoretically, the in-plane thermal transport along any direction $\alpha$ could be simplified as the results of two parallel paths along *x*- and *y*-directions (Figure S5d). For example, for the parallel-aligned CNT film, the in-plane thermal conductivity along the direction $\alpha$ could be written as

$$k_{0,\alpha} = A_x k_\perp m \frac{Bi_x}{1+Bi_x} sin^2\alpha + A_y k_\| nm \frac{Bi_y}{1+Bi_y} cos^2\alpha, \qquad (2)$$

where $A_x$ and $A_y$ are the cross-section areas of a single CNT bundle normal to the *x*- and *y*-axis direction, respectively, *n* is the density of CNT bundles in a layer (the number of bundles per unit length perpendicular to the alignment direction), *m* is the density of the CNT layers (the number of layers per unit length along the z-axis), and $k_\|$ is the thermal conductivities of CNT bundles parallel to the axial direction. $Bi_x$ and $Bi_y$ are defined as $Bi_x = h_x D/A_x k_\perp$ and $Bi_y = h_y L_0/A_y k_\|$, respectively. $h_x$ and $h_y$ are the contact thermal conductance of CNT bundles along the *x*-axis and *y*-axis direction, respectively. More details of this theoretical model can be found in the Supporting Information. Equation 2 suggests that in the direction $\alpha = 0°$, thermal transport is determined by the axial thermal resistance of the CNT and the contact resistance between CNT bundles. Due to the weak van der Waals interfacial bonds, the thermal conductivity along the aligned direction ($\alpha = 0°$) was reduced to 1/10 (~100 W/mK) in spite of the ultra-high thermal conductivity of a single CNT (~ 1000 W/mK).[28] The thermal conductivity is reduced even more in the direction $\alpha = 90°$ due to the dominant effect of contact resistance between neighboring tubes. Our analysis suggests that the highly anisotropic in-plane thermal conductivity of the parallel-aligned CNT film originates from the combined effect of the very high thermal conductivity of individual CNTs along the axial direction and the poor thermal contact between neighboring tubes. Similarly, the directional dependence of in-plane thermal conductivity of CNT films with stacking angle of $\theta=90°$ and $45°$, $k_{90,\alpha}$ and $k_{45,\alpha}$, were also obtained (Supporting Information, Equation. S34 and S37). The $k_r$ of super-aligned CNT films with different stacking angles predicted by the theoretical model are plotted as curves in Figure 4, which agrees qualitatively well with our measurements. Our results suggest that the alignment of nanotubes plays an important role in the anisotropic thermal transport of super-aligned CNT films. By taking advantage of the anisotropic nature of individual CNTs, the in-plane thermal transport of super-aligned CNT films could be tuned with the in-plane anisotropy ranging from 13.5 (ultra-anisotropic) to 1.2 (homogenous) by simply changing the stacking angle of the CNT layers.

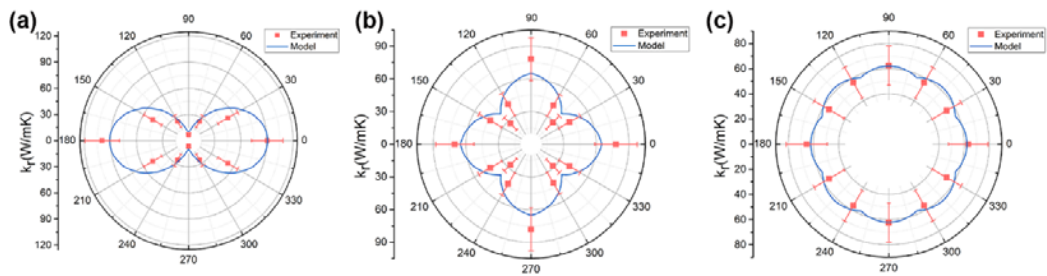

*Figure 4. The directional dependence of in-plane thermal conductivity for CNT films with stacking angle of (a) $\theta = 0°$, (b) $\theta = 90°$ and (c) $\theta = 45°$.*

**Conclusions**

In summary, the super-aligned CNT films with different alignment directions were designed to tune their thermal transport in both in-plane and cross-plane directions. The anisotropic thermal conductivities of CNT film were measured using the time-domain thermoreflectance (TDTR) measurement technique with the elliptical beam approach. To meet the requirement of an optically flat and reflective surface in the TDTR method, a special sandwich sample configuration was designed. It is found that the cross-plane thermal conductivity could be tuned from 6.4 to 1.5 W/mK when the stacking angle $\theta$ increases from 0° to 90°. The in-plane anisotropic ratio changed from 13.5 to 1.2 when the stacking angle changes from $\theta=0°$ to at $\theta=45°$, indicating that the in-plane thermal transport can be tuned from being highly anisotropic to homogenous. Analysis of our results suggests that the tunable anisotropic in-plane thermal conductivity originates from the ultrahigh axial thermal conductivity of a single CNT in contrast to the weak inter-tube thermal contact, together with the important role of the alignment. This work not only deepens our understanding of the influence of the CNT alignment on the anisotropic thermal transport in super-aligned CNT films but also provides a solution to measure the thermal conductivity of the porous films with high anisotropy.

**Supporting Information**

1. Verification of TDTR system
2. Sensitivity analysis for the cross-plane and in-plane thermal conductivity
3. Thermal model for the elliptical-beam method based on TDTR

4. Thermal conductivity Model of aligned CNT films

5. Preparation of random CNT film and its thermal conductivity

**Author Contributions**

The manuscript was written through the contributions of all authors. All authors have given approval to the final version of the manuscript. ‡These authors contributed equally.

**Corresponding Author**

* Email: ronggui@hust.edu.cn

chliu@mail.tsinghua.edu.cn

**Notes**

The authors declare no competing financial interest.

**Acknowledgments**

This work was supported by the National Key Research & Development Program of China (2018YFA0208401).

# References

1. Kordás, K.; Tóth, G.; Moilanen, P.; Kumpumäki, M.; Vähäkangas, J.; Uusimäki, A.; Vajtai, R.; Ajayan, P. M. *Applied Physics Letters* **2007,** 90, (12), 123105.
2. Teng, T.-P.; Teng, T.-C. *Applied Thermal Engineering* **2013,** 51, (1-2), 1098-1106.
3. Kaur, S.; Raravikar, N.; Helms, B. A.; Prasher, R.; Ogletree, D. F. *Nat Commun* **2014,** 5, 3082.

4. Hsu, P. C.; Liu, X.; Liu, C.; Xie, X.; Lee, H. R.; Welch, A. J.; Zhao, T.; Cui, Y. *Nano Lett* **2015,** 15, (1), 365-71.
5. Cui, Y.; Gong, H.; Wang, Y.; Li, D.; Bai, H. *Adv Mater* **2018,** 30, (14), e1706807.
6. Hone, J.; Llaguno, M. C.; Nemes, N. M.; Johnson, A. T.; Fischer, J. E.; Walters, D. A.; Casavant, M. J.; Schmidt, J.; Smalley, R. E. *Applied Physics Letters* **2000,** 77, (5), 666-668.
7. Zhang, K. J.; Yadav, A.; Kim, K. H.; Oh, Y.; Islam, M. F.; Uher, C.; Pipe, K. P. *Adv Mater* **2013,** 25, (21), 2926-31.
8. Zhao, X.; Huang, C.; Liu, Q.; Smalyukh, I. I.; Yang, R. *Journal of Applied Physics* **2018,** 123, (8), 085103.
9. Prasher, R. S.; Hu, X. J.; Chalopin, Y.; Mingo, N.; Lofgreen, K.; Volz, S.; Cleri, F.; Keblinski, P. *Phys Rev Lett* **2009,** 102, (10), 105901.
10. Jiang, K.; Wang, J.; Li, Q.; Liu, L.; Liu, C.; Fan, S. *Advanced Materials* **2011,** 23, (9), 1154-1161.
11. Jiang, S.; Liu, C.; Fan, S. *ACS Appl Mater Interfaces* **2014,** 6, (5), 3075-80.
12. Wang, D.; Song, P.; Liu, C.; Wu, W.; Fan, S. *Nanotechnology* **2008,** 19, (7), 075609.
13. Jiang, P.; Qian, X.; Yang, R. *Rev Sci Instrum* **2018,** 89, (9), 094902.
14. Jiang, P.; Qian, X.; Li, X.; Yang, R. *Applied Physics Letters* **2018,** 113, (23), 232105.
15. Zhang, L.; Zhang, G.; Liu, C.; Fan, S. *Nano Lett* **2012,** 12, (9), 4848-52.
16. Liu, K.; Sun, Y.; Chen, L.; Feng, C.; Feng, X.; Jiang, K.; Zhao, Y.; Fan, S. *Nano Letters* **2008,** 8, (2), 700-705.
17. Jiang, K.; Li, Q.; Fan, S. *Nature* **2002,** 419, (6909), 801-801.
18. Jiang, P.; Qian, X.; Yang, R. *Rev Sci Instrum* **2017,** 88, (7), 074901.
19. Cahill, D. G. *Review of Scientific Instruments* **2004,** 75, (12), 5119-5122.
20. Taphouse, J. H.; Smith, O. N. L.; Marder, S. R.; Cola, B. A. *Advanced Functional Materials* **2014,** 24, (4), 465-471.
21. Bougher, T. L.; Taphouse, J. H.; Cola, B. A., Characterization of Carbon Nanotube Forest Interfaces Using Time Domain Thermoreflectance. 2015.
22. Lin, W.; Zhang, R.; Moon, K.-S.; Wong, C. P. *Carbon* **2010,** 48, (1), 107-113.
23. Jiang, P.; Qian, X.; Yang, R. *Journal of Applied Physics* **2018,** 124, (16), 161103.
24. Hamby, D. M. *Environmental Monitoring and Assessment* **1994,** 32, (2), 135-154.
25. Hone, J.; Llaguno, M. C.; Biercuk, M. J.; Johnson, A. T.; Batlogg, B.; Benes, Z.; Fischer, J. E. *Applied Physics A* **2002,** 74, (3), 339-343.
26. Inoue, Y.; Suzuki, Y.; Minami, Y.; Muramatsu, J.; Shimamura, Y.; Suzuki, K.; Ghemes, A.; Okada, M.; Sakakibara, S.; Mimura, H.; Naito, K. *Carbon* **2011,** 49, (7), 2437-2443.
27. Schmidt, A. J.; Chen, X.; Chen, G. *Rev. Sci. Instrum.* **2008,** 79, (11), 114902.
28. Kim, P.; Shi, L.; Majumdar, A.; McEuen, P. L. *Phys Rev Lett* **2001,** 87, (21), 215502.